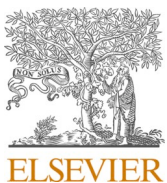
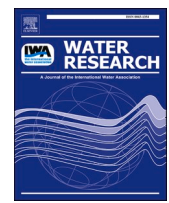

# LSTM networks provide efficient cyanobacterial blooms forecasting even with incomplete spatio-temporal data

Claudia Fournier [a], Raúl Fernandez-Fernandez [b], Samuel Cirés [a], José A. López-Orozco [b], Eva Besada-Portas [b], Antonio Quesada [a],*

[a] *Departamento de Biología, Universidad Autónoma de Madrid, 28049 Madrid, Spain*
[b] *Departamento de Arquitectura de Computadores y Automática, Universidad Complutense de Madrid, 28040 Madrid, Spain*

A B S T R A C T

Cyanobacteria are the most frequent dominant species of algal blooms in inland waters, threatening ecosystem function and water quality, especially when toxin-producing strains predominate. Enhanced by anthropogenic activities and global warming, cyanobacterial blooms are expected to increase in frequency and global distribution. Early Warning Systems (EWS) for cyanobacterial blooms development allow timely implementation of management measures, reducing the risks associated to these blooms. In this paper, we propose an effective EWS for cyanobacterial bloom forecasting, which uses 6 years of incomplete high-frequency spatio-temporal data from multiparametric probes, including phycocyanin (PC) fluorescence as a proxy for cyanobacteria. A probe agnostic and replicable method is proposed to pre-process the data and to generate time series specific for cyanobacterial bloom forecasting. Using these pre-processed data, six different non-site/species-specific predictive models were compared including the autoregressive and multivariate versions of Linear Regression, Random Forest, and Long-Term Short-Term (LSTM) neural networks. Results were analyzed for seven forecasting time horizons ranging from 4 to 28 days evaluated with a hybrid system that combined regression metrics (MSE, $R^2$, MAPE) for PC values, classification metrics (Accuracy, F1, Kappa) for a proposed alarm level of 10 µg PC/L, and a forecasting-specific metric to measure prediction improvement over the displaced signal (skill). The multivariate version of LSTM showed the best and most consistent results across all forecasting horizons and metrics, achieving accuracies of up to 90 % in predicting the proposed PC alarm level. Additionally, positive skill values indicated its outstanding effectiveness to forecast cyanobacterial blooms from 16 to 28 days in advance.

## 1. Introduction

Cyanobacteria can form dense blooms in aquatic systems. These blooms can lead to a marked discoloration of the water, interfere with human recreation activities, affect the quality of drinking water supply and fisheries, and produce secondary metabolites (e.g., cyanotoxins) that can be dangerous to human health (Huisman et al., 2018). In the last decades, increasing nutrient over-enrichment in aquatic systems due to anthropogenic activities, coupled with rising temperature due to global warming, has resulted in optimal conditions for an increase in cyanobacterial growth rates (Paerl and Otten, 2013). This fact suggests that cyanobacterial blooms will become more frequent over time and in a broader geographical distribution (Chapra et al., 2017; Paerl and Otten, 2013) leading to an economic impact of billions of dollars annually (Ho et al., 2019; Sanseverino et al., 2016). As a countermeasure, a range of preventive, control, and mitigation measures (such as in-lake nutrient control techniques or application of cyanocides) have been explored to restrict cyanobacterial growth to safe levels (Ibelings et al., 2016). To be able to implement these measures before the situation escalates, it is still necessary to develop effective Early Warning Systems (EWS) capable of predicting when the cyanobacterial bloom is expected to occur. As part of these EWS, multiple monitoring tools are available to detect the onset of cyanobacterial blooms. These approaches range from simple verifications such as human visual inspection to more sophisticated techniques such as remote sensing or molecular methods (Almuhtaram et al., 2021). In this context, fluorescence probes have gained popularity as they provide a reliable proxy for the cyanobacterial presence through chlorophyll *a* (CHLA) and phycocyanin (PC) estimates. These sensors have demonstrated a significant potential to be integrated in Artificial Intelligence

\* Corresponding author.
*E-mail address:* antonio.quesada@uam.es (A. Quesada).

(AI) models to forecast cyanobacteria levels (Bertone et al., 2018). Within this framework, Rousso et al. (2020) analyzed over 120 studies with forecasting and predictive models (including, among others, regressions, artificial neural networks, decision trees and probabilistic models) for cyanobacterial blooms across global freshwaters systems. Although most of these studies introduce in situ fluorescence measurements as one of the main data sources, this review pointed out an overall lack of replicability and comparability across studies since most of the reviewed models were site and species-specific, and conclusions regarding the main predictors were highly variable. Moreover, the results were typically obtained with a single model, making it difficult to compare the performance of different models due to the high variability introduced by the diverse datasets used in each publication. In addition to this problem, in the current state of the art for cyanobacterial bloom forecasting and predictive models, there seems to be a consensus on the use of CHLA as a proxy for cyanobacterial blooms (Chen et al., 2015; Mozo et al., 2022). However, several studies have raised skepticism about whether the use of CHLA is appropriate for an accurate cyanobacterial monitoring as this pigment is also present in other phytoplanktonic groups (Almuhtaram et al., 2021a). To address this issue, other parameters, such as PC fluorescence or the number of cyanobacterial cells per milliliter, have been recently suggested to be more appropriate for cyanobacterial bloom forecasting (Ahn et al., 2023; Almuhtaram et al., 2021). In addition, only some of the studies in Rousso et al. (2020) focus on exploring short-term forecasts, and only a limited number of them assess the performance of the models at different forecasting horizons (Thomas et al., 2018). Another relevant aspect to consider is the spatial distribution of cyanobacteria and the ability of some bloom-forming genera to move vertically in the water column through controlled buoyancy (D'Alelio et al., 2011). Most studies use only fluorometric measurements derived from probes fixed at the water surface, which have no capability to capture cyanobacterial development at depths different from the surface (Almuhtaram et al., 2021b; Mozo et al., 2022). However, using data derived from profilers often results in incomplete spatio-temporal data, a common problem with automated sensors when considering the integration of these data into forecasting models. Hence, monitoring tools (such as fluorometric probes mounted on water profilers) and forecasting techniques that account for this variability should be strongly considered. Finally, it is worth noting that although there are different options to assess the performance of forecasting models, only two of them have been widely applied: the first is quantitative assessment through regressive metrics to assess the capability of the models to forecast the exact value of a given parameter (Ahn et al., 2023); while the second is qualitative assessment through classification metrics to assess the capability of the models to forecast whether the value of a parameter will exceed a predefined threshold, such as an alert level (Park et al., 2021). However, since traditional metrics have limited capabilities in forecasting tasks (Alexander et al., 2015; Hyndman and Koehler, 2006), several authors highlight the importance for a more comprehensive assessment combining assemblies of traditional metrics (Sokolova and Lapalme, 2009) with a more holistic approach based on a critical discussion of the results (Saffo, 2007; Zhu et al., 2023).

The main aim of this paper is to propose an effective EWS for forecasting cyanobacterial blooms which overcomes the limitations of existing models. As previously mentioned, current forecasting models are often site/species-specific, rely on incomplete surface measurements, and primarily use CHLA as a proxy, which may not accurately represent cyanobacterial concentrations. Additionally, there is a lack of comparability across studies, as models are rarely trained on the same dataset or evaluated over different forecasting horizons. In this context, the presented approach introduces several novel elements:

- It is supported by non-site/species-specific models trained on data obtained from a multiparametric probe mounted on a water column profiler instead of relying solely on surface measurements, allowing for the monitoring of vertical distribution patterns and their temporal variability.
- It employs biologically guided pre-processing of incomplete high-frequency data using PC as the bloom proxy to ensure flexibility and replicability with data from other probes.
- It includes a comparison of six different models, each evaluated using the same dataset. These include classical regressive models, ensemble machine learning models, and state-of-the-art deep learning models, each in their autoregressive and multivariate versions.
- It is evaluated through a hybrid method, which combines traditional quantitative metrics with a qualitative alert level system and a holistic skill metric, applied over multiple temporal forecasting horizons. This approach allows for the assessment of the models under varying environmental conditions, enhancing their interpretability and applicability in real-world scenarios.

By addressing the limitations of current models, this paper provides a versatile option for effective management, adding increased time for the implementation of management measures to reduce the risk associated with cyanobacterial blooms.

## 2. Materials and methods

The workflow followed in this paper is presented in Scheme 1 and further explained in the following subsections.

### 2.1. Site description

Cuerda del Pozo is a multipurpose reservoir located at the headwaters of the Duero River in Soria, Spain (Fig. 1). This water resource serves as a public water supply, but it is also used for recreation, irrigation, and hydroelectric power generation. It is eutrophic and presents a monomictic regime, covers an area of 2289 ha, and has a capacity of 249 $hm^3$ with average and maximum depths of 10 and 36 m, respectively.

### 2.2. Data collection

Aiming to understand cyanobacterial dynamics and water quality interactions in Cuerda del Pozo reservoir, Ecohydros S.L. (Cantabria, Spain) installed an automated multiparametric probe mounted on a floating platform (Fig. 1) near the reservoir dam. The monitoring took place over the course of six years, from January 2010 to November 2015, with sensors routinely calibrated and overall functionality checked. The probe was specifically designed to capture daily data throughout the entire water column with an automatic profiler. Moreover, multiple values were taken hourly within the photic zone, typically ranging from 2 to 4 m, resulting in a dataset of >50,000 readings acquired at different depths. The acquisition frequency of these data was not constant, due to technical issues, leading to periods of missing values, and spatial and temporal incompleteness in the collected data. The parameters considered for this paper include fluorometric measurements, taken simultaneously at the same depth, of PC and CHLA obtained using a TriOS probe (microFlu-blue and microFlu-chl sensors), as well as regular water temperature measurements (TEMP) using the CTD60M probe (Sea & Sun Technology). Further specifications about the platform, probes and sensors are provided in detail in the work of Monteoliva (2016). Although cyanobacterial blooms are complex events enhanced by the interaction of biological, physicochemical, and hydromorphological factors (Huisman et al., 2018), the reason for selecting only these variables was twofold. First, the main objective of this work is to obtain optimal models for the prediction task (rather than describing influences between parameters and cyanobacterial blooms), for which it is recommended to use a few key parameters, maximizing the feature-to-size ratio (Zhu et al., 2023). Second, the proposed methodology is intended to serve as a guide for other users of multiparametric probes, and therefore the selection of a few parameters that are easy to measure in an automated manner is crucial. In this way, CHLA and PC represented the variables automatically collected with the highest correlation to phytoplanktonic and cyanobacterial biomass. In addition, TEMP represented one of the most remarkable variables influencing cyanobacterial blooms (Paerl et al., 2008). This variable is assumed to have great capacity to integrate information on how other variables affect the water regime (e.g., residence time and precipitation can affect mixing regimes or stratification patterns, which, in turn, can impact and be reflected in the temperature throughout the water column), making it a good parameter to consider.

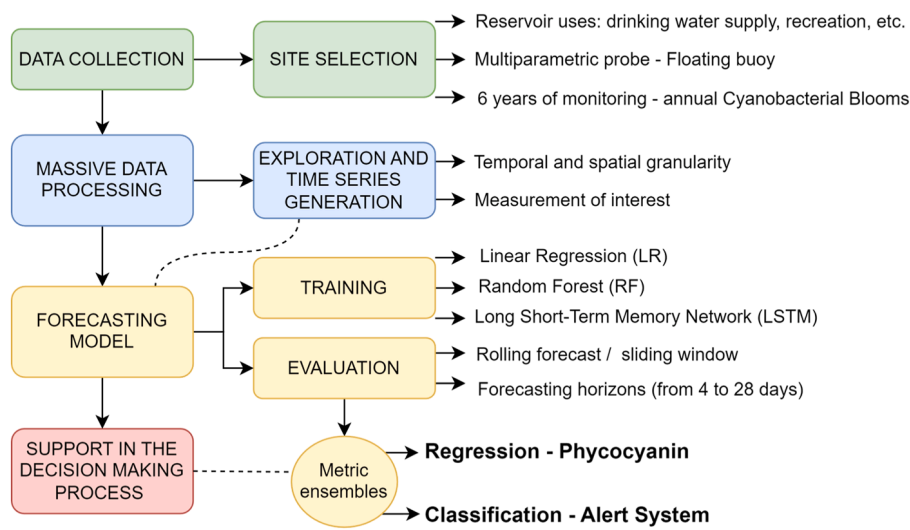

**Scheme 1.** Workflow followed in this paper.

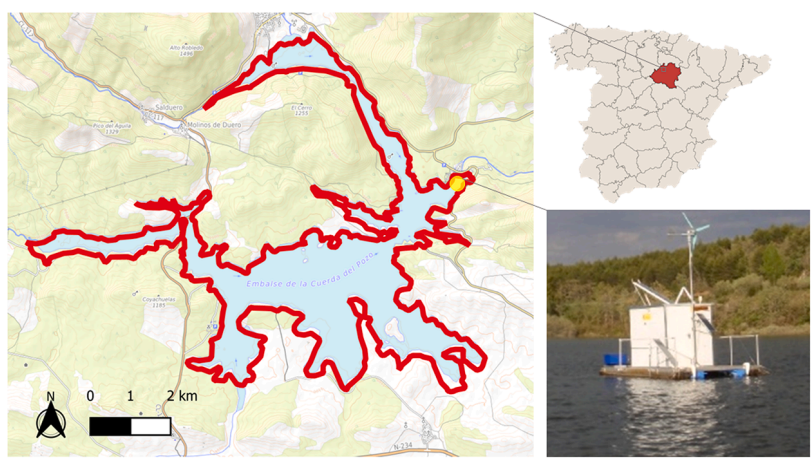

**Fig. 1.** Location of Cuerda del Pozo reservoir (Soria, Spain) and the floating platform.

During the 6 years of sampling, multiple cyanobacterial blooms were found (see Fig. S1 in supplementary material) taking place mainly in the months of September and October. Two of these blooms (October 2010 and October 2013) were notably remarkable, registering unusually high levels of PC fluorescence and being dominated by potentially toxic and bloom-forming cyanobacterial genera, such as *Dolichospermum, Aphanizomenon*, and *Microcystis* (Monteoliva, 2016).

*2.3. Data exploration*

To enhance the effectiveness of PC concentration forecasting in the context of cyanobacterial blooms, the initial step involved an in-depth data exploration aimed at facilitating a biologically guided pre-processing of the original dataset. Conventional descriptive statistical techniques (such as the mean, standard deviation, quantiles) and graphical representations (like histograms and box plots) were used to analyze the distribution, centrality, and dispersion of each parameter. A multivariate analysis, focused on Spearman correlations and time series graphical representations, was performed to study the relationship between parameters and their spatio-temporal patterns. In this process, periods and depths with missing values were also identified, along with anomalous values or outliers.

*2.4. Data pre-processing*

In this phase, multiple steps were carried out, in order to pre-process the data into a time series dataset optimized for cyanobacterial bloom forecasting. The objectives were twofold. The first one was to construct different time series for each parameter using information regarding maximum values (across the entire water column, not limited to the photic zone) and integrated values (averaging all data points within the water column). The second objective was to make this process easily automated and replicable with data from other probes, even if their operational characteristics were different.

To achieve these objectives, the following steps were performed to explore the temporal (time) and spatial (depth) distributions, as well as the values of each parameter considered:

(1) Determining the optimal temporal granularity to reduce high frequency noise, without excessive data smoothing or relevant information loss. Daily, 4-day, and weekly intervals were considered.
(2) Identifying the best statistical measures that fit the distribution of maximum and integrated values while minimizing sensitivity to potential outliers. Measures such as mean, median, third quartile (Q3), maximum, and the total sum were explored.
(3) Determining the optimal depth granularity to reduce measurement errors without relevant information loss. Three different scenarios were considered: the complete water column, the initial 20 m from the surface, and the initial 10 m from the surface.

Afterwards, both the time series for maximum and the time series for integrated values were generated excluding the first meter due to excessive noise introduction. For integrated values, the heterogeneity distribution of measurements across depths, such as only sampling within the euphotic zone on certain dates, led to biased estimates. To address this problem, a 2D linear interpolation was first performed considering both time and depth simultaneously, which provided depth homogeneity across all days within the studied interval. The resulting interpolated dataset facilitated the generation of integrated time series. In contrast, time series of maximum values were obtained without prior interpolation just selecting the most appropriate statistical measure (among the mean, median, Q3, and maximum values). As a result, time series with missing values was obtained and interpolated afterwards using various methods found in literature for the modelling of environmental data (Jasiński, 2015). Different interpolation methods were explored to identify the approach that was better aligned with the natural behavior of each parameter. Ultimately, spline interpolation was used for all parameters, except for CHLA which was interpolated linearly. Finally, all resulting time series, including both maximum and integrated values, were compared. Time series corresponding to highly correlated features were discarded, maximizing the feature-size ratio (SFR) for integration into the forecasting models.

*2.5. Forecasting models – Setup and training*

Each time series was divided into two identical portions (50/50, before and after the 1st of January 2013) in order to construct the training and test datasets. While it is a good practice to split the data around 60/40 or 70/30 for training and testing, respectively (Zhu et al., 2023), the 50/50 split was chosen to include, within the testing set, the blooming period of October 2013, where high levels of PC fluorescence were observed. Hence, the other relevant blooming period (occurred in October 2010) was included in the training dataset.

The autoregressive and multivariate versions of three types of time series forecasting models, ranging from classic to state-of-the-art models, were selected and compared in terms of interpretability and performance. A comparison was performed to determine whether the increased complexity of newer models, such as Long Short-Term Memory (LSTM) networks, is justified by improving the predictive capability with respect streamlined models, such as Linear Regressions (LR), and models of intermediate complexity, such as the Random Forest (RF). To ensure comparability, all models had access to the same dataset and variables, considering a maximum of four months of data as input, for making predictions. Different configurations were tested to determine the optimal time periods of the input variables and the selection of variables to be considered for each model. Finally, the most effective configurations were selected.

On one hand, LR was selected as an effective model for data exhibiting linear relationships, which has already showed promising results in forecasting harmful algal blooms (Chen et al., 2015). It is a technique that forecasts future values of a variable by modelling its linear relationships with its past observations (autoregressive) and one or more independent/exogenous variables (multivariate) as shown in Eq. (1) and Eq. (2), respectively.

$$\widehat{y}_t = \beta_0 + \beta_1 y_{t-1} + \beta_2 y_{t-2} + \ldots + \beta_p y_{t-p} + \epsilon_t \quad (1)$$

$$\widehat{y}_t = \beta_0 + \beta_1 y_{t-1} + \beta_2 y_{t-2} + \ldots + \beta_p y_{t-p} + \gamma_1 x_{1,t} + \gamma_2 x_{2,t} + \ldots + \gamma_q x_{q,t} + \epsilon_t \quad (2)$$

where $\widehat{y}_t$ is the predicted value, $y_{t-1}, y_{t-2}, \ldots, y_{t-p}$ are the past values of the predicted variable, $\beta_1, \beta_2, \ldots, \beta_p$ are their coefficients, $\beta_0$ is the y-intercept, $\epsilon_t$ is the error at time $t$, $x_{1,t}, x_{2,t}, \ldots, x_{q,t}$ are the additional variables at time $t$, and $\gamma_1, \gamma_2, \ldots, \gamma_p$ are their coefficients.

Only PC was considered in the autoregressive model, and both CHLA and TEMP were incorporated in the multivariate version. A scaling encoder was applied to normalize all variables (scaled between 0 and 1), and randomness was set to 42 to ensure reproducibility across experiments. Only data from the 12 previous days was used as input, as short periods yielded the best results. Both LR models were implemented using the Darts (v0.26.0) forecasting library in Python.

On the other hand, the RF algorithm was chosen due to its outstanding performance in previous forecasting tasks in literature when compared to other well-established machine learning models (Dudek, 2022; Harris and Graham, 2017; Kane et al., 2014). This algorithm is designed to work with noisy data and outliers, while having the capability to identify complex non-linear relationships. It is an ensemble learning technique that constructs multiple decision trees, each based on a random selection of samples and parameters, enhancing model performance and robustness (Breiman, 2001). The final prediction is determined by averaging the outputs of all individual trees. Both the autoregressive and multivariate RF were trained to make predictions using PC data from the previous 12 days. The multivariate RF model also incorporated CHLA and TEMP values, as this configuration yielded the best results. As for LR, a scaling encoder was applied to normalize all variables (scaled between 0 and 1), and randomness was set to 42 to ensure reproducibility across experiments.. Both models comprised 100 trees with no maximum depth restriction. The remaining parameters were set to their default values. Both RF models were implemented using the Darts (v0.26.0) forecasting library in Python.

Finally, the LSTM Networks (Hochreiter and Schmidhuber, 1997), which belong to the group of Recurrent Neural Networks (Bengio et al., 1994), were selected as a method to introduce the most recent advances of time series forecasting using deep learning given its demonstrated success in recent research and publications within this field (Freeman et al., 2018; Gajamannage et al., 2023; Karevan and Suykens, 2020). The main difference is the presence of loops within layers, so past information can be used for the inference of new one, providing capability to store information for higher numbers of time steps. This feature is particularly beneficial for forecasting cyanobacterial blooms, as these often exhibit complex temporal dependencies and long-term patterns that can be effectively captured by LSTM architecture. As with previous models, one autoregressive and one multivariate model were defined. The multivariate model introduced TEMP values as the only exogenous variable. The parameter CHLA was not included because it reduced the performance of the models when used as input during pilot experiments. Data from the last 4 months was used as input for this model. Using this input, an additional data augmentation step was introduced with the code proposed by Iwana et. al (2021). Training was performed until reaching five consecutive epochs without improving the validation loss. The training setup included a learning rate of 0.0001, mean squared error as the function loss, and a batch size of 5. The rest of the training setup is defined within the supplementary materials in Table S2. Both LSTM models were implemented using TensorFlow with GPU support (version 2.4.1) and Keras (version 2.4.3) in Python.

*2.6. Forecasting models – Evaluation of performance*

The performance of the models was assessed on the testing set (see

**Table 1**
Evaluation metrics: name, formula and brief definition.

| | Metric | Formula | Definition |
|---|---|---|---|
| Regression metrics – PC concentration | MSE | $\frac{1}{n_{samples}} \sum_{i=0}^{n-1} (y_i - \hat{y}_i)^2$ | Average of the squared differences between predicted and actual values. Lower values indicate better performance. |
| | $R^2$ | $1 - \frac{\sum_{i=1}^{n} (y_i - \hat{y}_i)^2}{\sum_{i=1}^{n} (y_i - \bar{y})^2}$ | Proportion of the variance in the dependent variable predictable from the independent variable(s). It ranges from 0 to 1, where 1 indicates a perfect fit and 0 indicates no linear relationship. |
| | MAPE | $\frac{1}{n_{samples}} \sum_{i=0}^{n-1} \frac{|y_i - \hat{y}_i|}{max(\epsilon, |y_i|)}$ | Percentage difference between predicted and actual values. Lower percentage indicates better performance. |
| | SS | $1 - \frac{MAPE_{forecast}}{MAPE_{displaced}} \cdot 100$ | Improvement percentage of forecasts over replicating the last observed value (displaced signal). Positive and higher values indicate better performance. |
| Classification metrics – alarm level | Acc | $\frac{(TP + TN)}{(TP + FP + TN + FN)}$ | Proportion of correctly classified instances. Values range from 0 to 1, where higher values indicate better performance. |
| | F1 | $2 \times \frac{(precision \times recall)}{(precision + recall)}$ | Balance between precision (TP / (TP + FP)), which is the ratio of true positives to predicted positives) and recall (TP / (TP + FN), which is the ratio of true positives to actual positives). |
| | Kappa | $\frac{\Pr(a) - \Pr(e)}{1 - \Pr(e)}$ | Agreement between observed and predicted instances considering the probability of random agreement. Values range from −1 to 1, higher values indicate better-than-chance agreement. |

In regression metrics $n$ represents the number of samples, $y_i$ the observed value, and $\hat{y}_i$ the predicted value. In classification metrics TP and TN are true positives and true negatives, respectively; FP and FN are false positives and false negatives, respectively; and Pr($a$) is the probability of observed agreement and Pr($e$) is the probability of random agreement.

Section 2.4). All models were evaluated on all the proposed forecasting horizons, spanning from less than one week to one month ahead. These horizons were chosen based on the optimal temporal granularity determined during the exploration phase. The evaluation was conducted using a rolling forecast approach (Liu et al., 2021) where both past and present observations, along with the forecasting horizon, were considered within a moving window to make predictions for the following period.

Performance was evaluated in a two-step approach. First, a visual inspection of the displayed predictions provided a qualitative assessment. Beyond the general fit of the predictions, this inspection focused on the fit to the PC concentration peaks following the Donaldson et al. (2023) approximation. Thus, the visual inspection considered how well the magnitude, timing, and shape of the relevant peaks (blooms in the context of this paper) were predicted in the time series forecasts. Magnitude understood as the height of the PC value at the peak (e.g., whether it was predicted to be higher or lower); timing as the exact moment when the peak occurred (e.g., whether the peak was predicted to occur earlier or later); and shape as the overall form of the rising and falling curve (e.g., whether the curve was predicted to be wider or narrower). Afterwards, the regression and classification metrics, presented in the following subsections, were computed providing a quantitative evaluation to be able to capture different aspects of the forecasting models (Sokolova and Lapalme, 2009; Zhu et al., 2023).

### 2.6.1. Regression metrics for PC forecasting

Bloom forecasting literature introduces traditional regression evaluation metrics to evaluate the performance of the proposed models. These metrics include Mean Absolute Percentage Error (MAPE), Mean Squared Error (MSE), and the coefficient of determination $R^2$ (Hyndman and Koehler, 2006). In addition, we propose the introduction of the forecast Skill Score (SS, Prado-Rujas et al., 2021) a specific forecasting metric that measures the performance of a forecast comparing it to a basic historical baseline of past observations. This baseline is defined as the real-time series displaced in the future by the same duration as the forecast horizon. The SS represents the capability of the forecast to outperform the simple replication of the last observed value. The SS value is computed using a regression evaluation metric, the forecasted time series and the displaced time series. In particular, in this paper MAPE is used for computing the SS value. Positive and higher values of SS indicate improvements of the forecasts when compared to the displaced signal. The mathematical expressions of all these metrics are displayed in Table 1.

### 2.6.2. Classification metrics for alert level forecasting

An alert system, built upon the regression models, was established to determine alert levels at PC concentrations equal to or above 10 μg/L based on both biological and data-driven considerations. From a biological standpoint, the chosen threshold aligns with a rational midpoint of Alert Level 1 from the World Health Organization (Chorus and Welker, 2021; Chorus and Bartram, 1999). The selected PC concentration corresponds to 10,000 cells/mL based on the analysis of Bastien et al. (2011) correlating TriOS probe readings with cyanobacterial cells per milliliter. Moreover, this value is consistent with others proposed in the literature that suggest alternative warning systems based on PC fluorescence (Ahn et al., 2007). Acknowledging the qualitative nature inherent in the interpretation of PC fluorescence measures and given their sensitivity to factors such as environmental interferences and fluorescence measurement limitations (Bertone et al., 2018), this selection supposes a rational and relatively flexible threshold within a conservative alert level. Simultaneously, from a data-driven perspective, the threshold of 10 μg/L was chosen as a best practice to maintain dataset balance allowing for an accurate evaluation of the forecast capability of the models by ensuring sufficient representation of both positive and negative cases (Shin et al., 2021; Zhu et al., 2023). Hence, Accuracy (Acc), F1 score (F1), and Cohen's Kappa coefficient (Kappa) were analyzed and computed, as stated in Table 1, to assess the capability of the models to effectively forecast the established alert level.

All analyses were conducted using in-house Python scripts (v3.9.2) executed within the Jupyter Notebook environment (Perkel, 2018). The python libraries used in this paper can be consulted in Table S3 of the supplementary material.

## 3. Results

### 3.1. Parameter distribution analysis

The univariate exploration of the depth, TEMP, CHLA and PC, without their temporal context, can be consulted in the supplementary material S4. This material includes the descriptive numerical analysis with statistical measurements (Table S4-A) and the distribution exploration of each variable (Fig. S4-B). This univariate exploration shows how the depth histogram was consistent with the expected performance of the probe with most data acquired in the photic zone and multiple measurements throughout the water column. In particular, PC exhibited a skewed right distribution, with an average of 5.5 μg/L, few maximum values exceeding 50 μg/L, and occasional values over 20 and 40 μg/L. PC levels above 10 μg/L were identified as extreme values, being 3.2 and 5.8 µg/L the first and third quartile values, respectively. CHLA followed a similar pattern with an average of 3.4 µg/L and maximum values above 30 µg/L. TEMP displayed a particular distribution with three predominant ranges, a mean of 12 ºC, and a minimum and maximum around 1 ºC and 25 ºC, respectively. The temporal evolution of each parameter exhibited some seasonality as shown in Fig. 2. The clearest seasonality was observed in the TEMP data as expected. Values started increasing in spring until the end of summer, when maximum TEMP gradient is observed. CHLA and PC levels tended to experience abrupt increases during the spring and autumn months, respectively, associated to blooming events. However, the magnitude and timing of these increases were highly variable, as well as the position of maximum values through the water column. Spearman correlation coefficients among parameters were low (< 0.27) (Fig. S5 in supplementary material).

### 3.2. Big data processing and time-series generation

The following section presents the results of the exploration into the best temporal and spatial granularity, as well as the selection of the best statistical measure for generating effective time series for cyanobacterial bloom forecasting. A graphical representation of some examples can be found in supplementary material S6. The temporal granularity was finally set to a 4-day interval for both the maximum and integrated dataset. This granularity was chosen to minimize the number of gaps in the data without compromising relevant information through excessive data smoothing. This interval is defined above the maximum time span necessary for an effective early warning, taking into account the division time of cyanobacteria (Chorus and Welker, 2021; Chorus and Bartram, 1999). Regarding statistical measures, the time series of maximum values was defined using the maximum value of each period. The reasons were that the percentile 95 and above values were able to accurately depict the behavior of the reservoir during the studied period and the absence of significant outliers. In datasets with significant outliers, using P95 instead of maximum values could mitigate their impact on the final time series. In the case of the time series of integrated values, the mean was introduced to define a statistical measure that takes into account all the values of the water column. For the selection of depth granularity, when considering the entire water column, a smoothing effect was observed for both series. Restricting the selection depth to values near the surface counteracted this smoothing effect making lower statistical position measures, such as percentile 75 or the median, effective for constructing time series of maximum values. However, it is crucial at this stage to observe the depth at which maximum values occur as certain cyanobacterial genera tend to accumulate closer to the thermocline whose position can vary over time and should not be excluded from the analysis. For this reason, the entire water column was considered for the generation of both time series, the one of maximum values and the one of integrated values.

In brief, the final time series of maximum values consisted of the maximum values within each 4-day interval considering data from the entire water column (Fig. 3), while the time series of integrated values consisted of mean values within each 4-day interval considering depths up to 24 m (Fig. 2). Since the time series of integrated values incorporated 2-D information, it was notably affected by numerous missing values during the first 3 years (Fig. 2). Nonetheless, both series exhibited a similar overall shape, although the latter displayed significant smoothing of information during critical periods (Fig. S7-A of the supplementary materials). Fig. 4 illustrates the correlations between the final time series (the ones of maximum and integrated values). Variability between parameters was consistently maintained as demonstrated by the low correlation coefficients. When comparing the correlation between the integrated and maximum time series, correlation coefficients for each parameter were high, ranging from 0.5 to 0.9. In addition, differences between integrated and maximum time series were mainly observed during periods of missing values (Fig. S7-A). For these reasons, the time series of integrated values were excluded from the subsequent analysis, which was conducted solely with the series of maximum values.

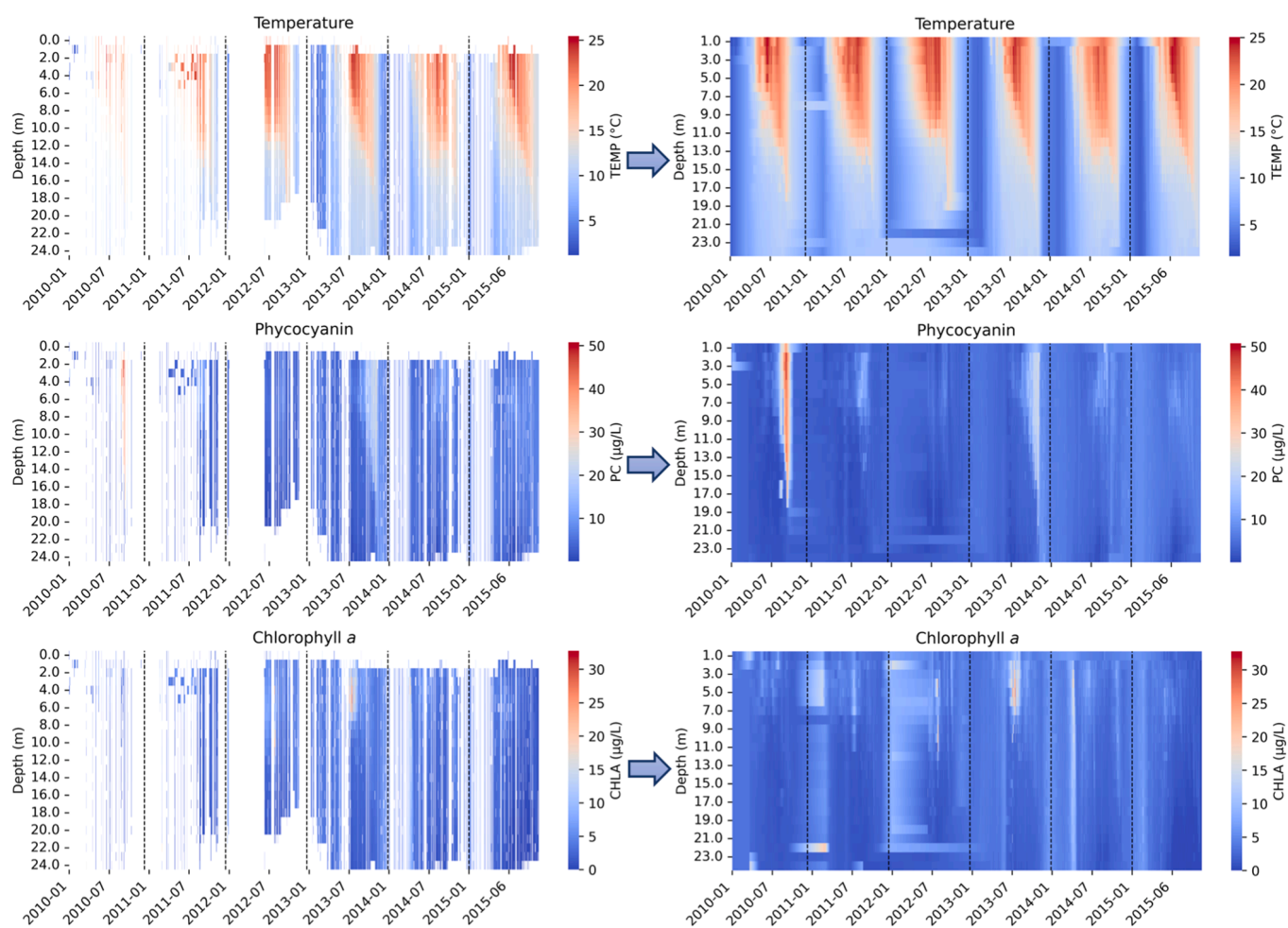

**Fig. 2.** Temporal (X-axis) and spatial (Y-axis) evolution of temperature (ºC), phycocyanin, and chlorophyll *a* (µg/L). Raw data (left) vs. 2D linearly interpolated data (right).

### 3.3. Models results and evaluation

Predictions for PC levels over different forecasting horizons for all models and versions (autoregressive and multivariate) are displayed in Fig. 5. A total of 6 models and 42 predictions are depicted in this figure (one for each of the seven forecasting horizons and for each model). For a detailed comparison between the autoregressive and multivariate versions of each model, see Figures in the supplementary material S8. In this section, first, a qualitative assessment by visual inspection of the results, focusing not just on the overall fit of the predictions but also on the timing, magnitude, and shape of PC peak predictions, is performed.

First the LR model exhibited predictions consistently close to real values. However, with increasing forecasting horizons, it becomes clear that its functionality was more reactive than predictive, showing a good fit in shape and magnitude but not in time. The same behavior is observed in its multivariate version. Hence, at short forecasting horizons, the LR model was good at predicting the magnitude and shape of the peaks at the cost of timing accuracy. In contrast, RF had the best performance in short forecasting horizons in relation to timing and magnitude of the most relevant PC peak within the first bloom period. This good performance, however, was not translated to the rest of the period introducing false positive peak predictions and relevant timing errors for other peaks. Incorporating exogenous covariates such as CHLA and TEMP improve RF predictions, which also tended to become more reactive than predictive beyond a 2–3 week horizon. Finally, the LSTM network had the best overall accuracy in terms of timing. This is the only model where all the predicted peaks occurred within the same bloom periods as the real peaks, while also maintaining a good shape performance. This good performance came at the cost of reducing magnitude accuracy to levels where EWS with high PC thresholds could be affected. However, the integration of TEMP data in the multivariate model strongly counteracted this deficit with no signs of overreaction at any forecast horizon.

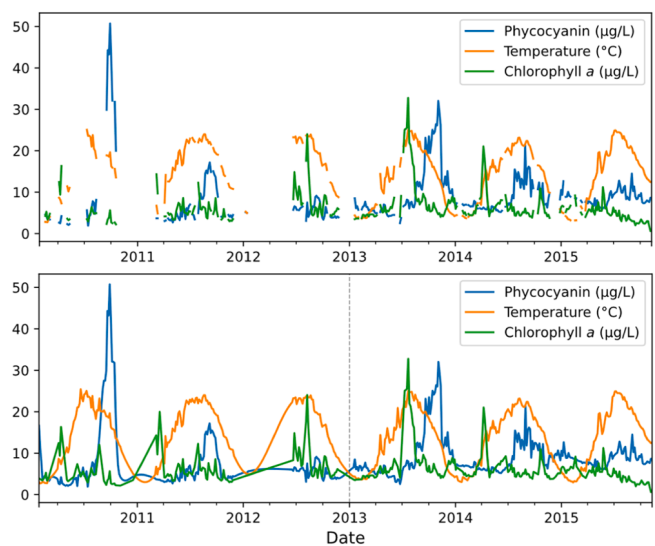

**Fig. 3.** Time series of maximum values of temperature (ºC), phycocyanin, and chlorophyll *a* (µg/L) with a time step of four days, before (above) and after (below) interpolation. Dashed vertical line marking the splitting date for training and testing sets.

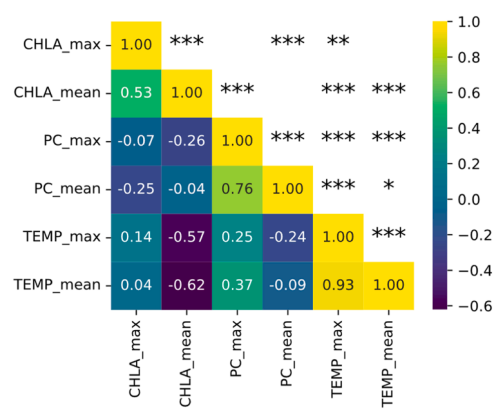

**Fig. 4.** Spearman correlation among time series of maximum and integrated/mean values. *, ** and *** for *p*-values < 0.05, 0.01 and 0.001, respectively. Circles denote pairs of values compared for maximum and integrated correlations.

Results of regression metrics, used to assess the capability of the models to predict PC values across different forecasting horizons, are shown in Fig. 6. Overall, a consistent pattern of loss in performance with higher horizons was observed across all metrics. Regarding error metrics (such as MSE and MAPE), LR models had better (lower) values at short forecast horizons, followed by the multivariate RF and autoregressive LSTM, and finally autoregressive RF and the multivariate LSTM. This trend rapidly shifted as forecasting horizons extended, with RF models, LR, and autoregressive LSTM experiencing an increase in prediction errors. Notably, the multivariate LSTM demonstrated consistency, with stable MSE and MAPE across forecasting horizons, standing out as the only model that was able to achieve this robust behavior. Meanwhile, the coefficient of determination ($R^2$, a fit metric where better is higher) also indicated a decline in model fitness when extending the forecasting horizons. Once again, the multivariate LSTM was the only model with consistent $R^2$ across forecasting horizons, indicating a maintained fit over time. Finally, the skill of every model increased with longer forecasting horizons, except for the LR models, whose skill remained relatively constant and close to zero, likely due to its reactive performance closely mirroring the displaced signal. This tendency shows the increased capability of every model to outperform the replication of the last observed value as the forecasting horizon extends. For short forecasting horizons (4–16 days) the autoregressive version of LR was the model with the highest skill values, although it did not outperform a purely displaced signal (as it had negative skill). After the 16–20 day horizon, the LSTM models were also able to achieve a positive skill value. These improvements went up to a maximum of a 25 % skill respect to the displaced signal for the LSTM model with exogenous variables. Neither of the two proposed RF variants demonstrated positive skill at any forecasting horizon.

Finally, Fig. 7 shows the evolution of classification metrics. These metrics are used to assess the capability of the models to predict an alarm level of PC values greater than 10 µg/L across different forecasting horizons. The overall accuracy of predictions ranged between 0.7 and 0.9. This means that accurate alarm predictions were achievable in 70–90 % of the cases for all models. However, as in the scenario for the PC level prediction, a discernible trend of loss in accuracy emerged as forecasting horizons increased. The autoregressive versions of all models experienced the most significant decrease in accuracy, especially in the case of RF. The multivariate version of all models maintained accuracies close to and above 80 %, even at long forecasting horizons. In particular, the multivariate version of LSTM was able to maintain accuracies around 90 % for all horizons. The same pattern became evident in the evolution of F1 score (a balance between precision and recall) and Kappa (a measure of model agreement beyond random chance). All models experienced substantial decreases in these metrics as forecasting horizons increased, with the multivariate LSTM being again the exception, showing stability and even some improvement over time. The autoregressive LSTM model, likely due to the observed loss of accuracy in magnitude (Fig. 5), showed the most pronounced impact at longer forecasting horizons followed by the autoregressive version of RF and LR. Finally, F1 score and Kappa from the multivariate versions of LR and RF did also decrease, both with a comparable performance, across forecasting horizons.

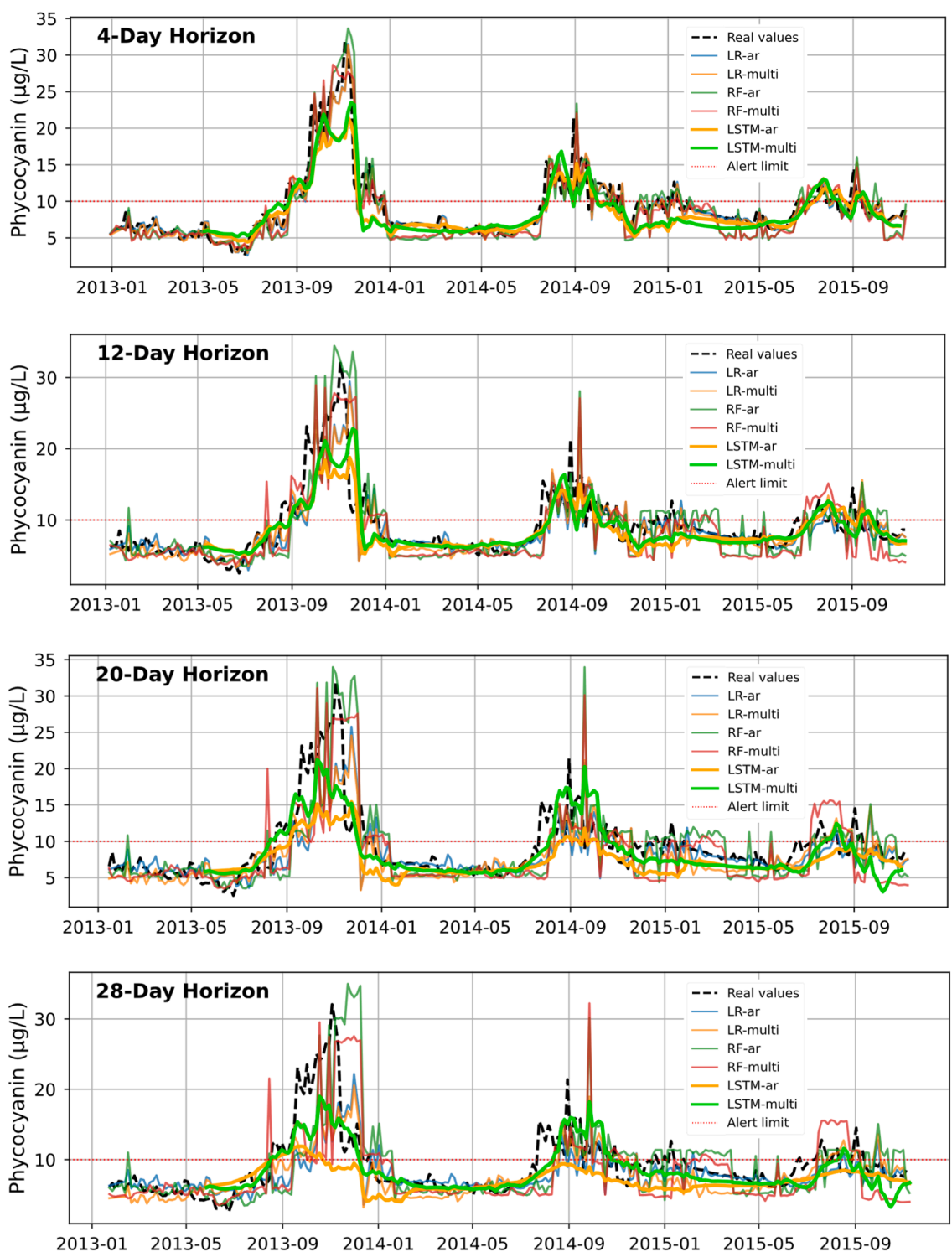

**Fig. 5.** PC predictions for the different forecasting horizons and models. The autoregressive and multivariate LR (LR-ar, LR-multi), RF (RF-ar, RF-multi), and LSTM neural network (LSTM-ar, LSTM-multi). Real PC values (black dashed line) are compared against predicted values. The alert limit at 10 µg/L of PC is represented by the red dashed line. Predictions for 8, 16, and 24 days ahead can be found in supplementary material S9.

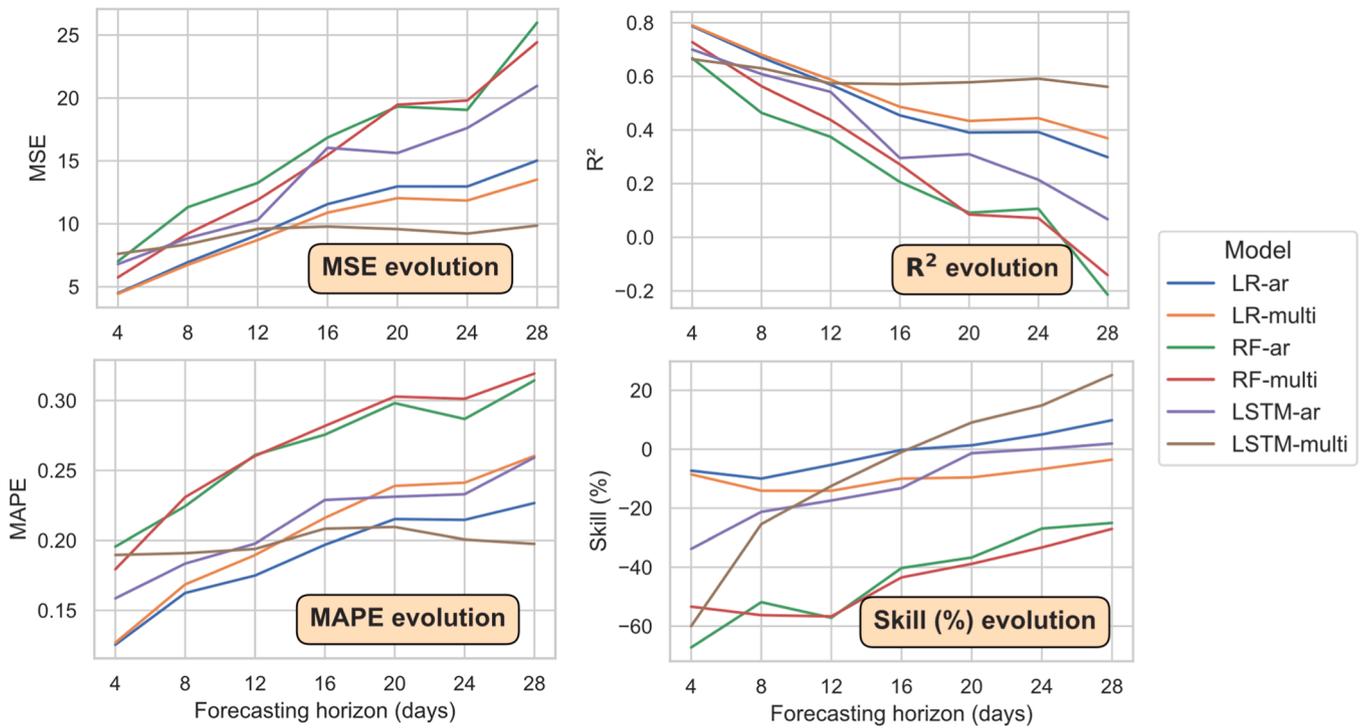

**Fig. 6.** Evolution of model performance for PC predictions with different regression metrics. The autoregressive and multivariate LR (LR-ar, LR-multi), RF (RF-ar, RF-multi), and LSTM neural network (LSTM-ar, LSTM-multi). MSE: Mean squared error; $R^2$: determination coefficient; MAPE: mean absolute percentage error; Skill: performance of model compared to the displaced signals (higher is better).

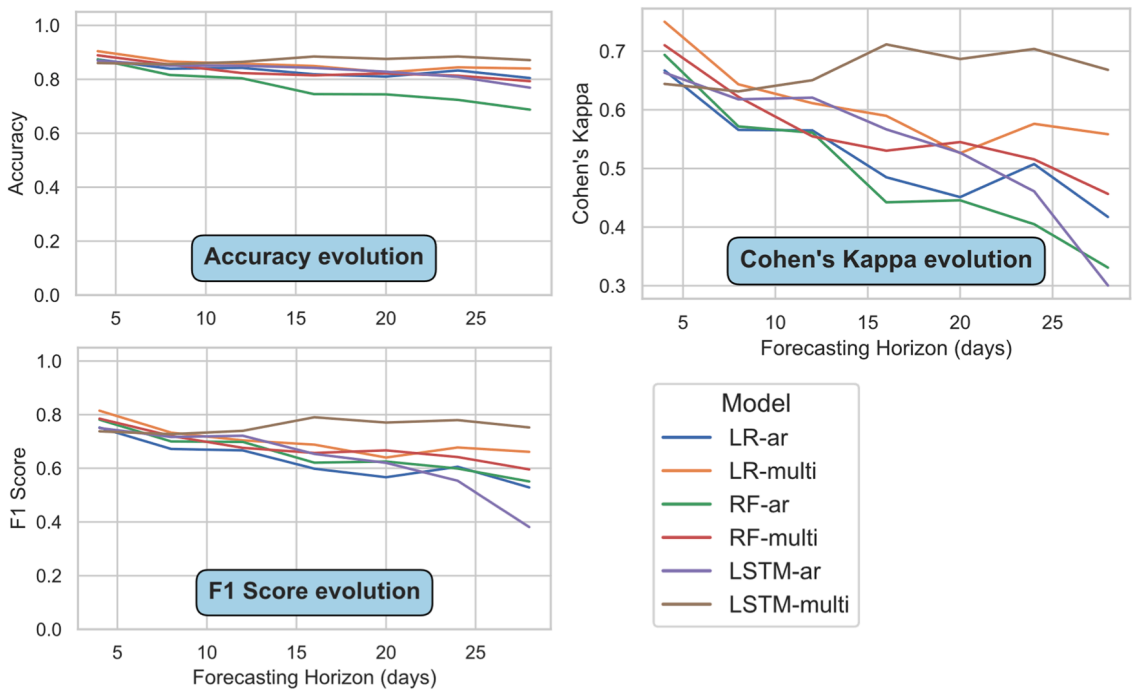

**Fig. 7.** Evolution of model performance for alarm level predictions with different classification metrics. The autoregressive and multivariate LR (LR-ar, LR-multi), RF (RF-ar, RF-multi), and LSTM neural network (LSTM-ar, LSTM-multi).

## 4. Discussion

In this paper an effective alert system is proposed to forecast PC concentrations between 4 and 28 days in advance. Data exploration and pre-processing played a key role in reaching meaningful time series for the forecasting of substantial increases in PC levels. Time and depth granularity, as well as the best statistical measurement, were effectively defined throughout the proposed biologically guided exploration. As this approach is flexible and easily replicable, users can adapt it to their specific datasets and objectives, shortening the gap between ecological forecasts and environmental decision making (Bodner et al., 2021). In addition, having access to the profiling data provided a comprehensive understanding of the cyanobacterial behavior and distribution in time and depth. This allowed us to answer the question of when an abrupt increase in PC concentration would occur, regardless of its specific location in the water column, greatly enhancing the scope of the results compared to those derived solely from surface data. In this specific case, the integrated values were excluded from the training of the forecasting models due to their high correlation with the time series of maximum values, their high amount of interpolated data, and the significant smoothing of information during critical periods. However, this may not be the case for other water systems or parameters. Therefore, it is highly recommended to perform a similar exploratory analysis before excluding integrated values from modelling in similar works. Regarding the selected variables, CHLA and PC values did not show a significant correlation in this paper, reinforcing the importance of the latter as a more specific proxy for cyanobacterial blooms (Ahn and Oh, 2007). Furthermore, TEMP showed great importance in improving model performance. These results support the hypothesis that the temperature throughout the water column is influenced by other variables that may affect cyanobacterial bloom development (e.g., hydrodynamic or environmental parameters such as water level, residence time, light hours or precipitation), while also reflecting critical aspects such as stratification regimes. Thus, TEMP may serve as an integrative variable that encapsulates the influence of various factors relevant to cyanobacterial bloom development, improving the forecasting capability of models. In future work, it is recommended to investigate the forecasting potential of other variables which can be measured with high spatio-temporal frequency as the ones studied in this paper. In order to test the ability of the presented models to be effectively integrated into EWS, an attempt has been made to relate PC fluorescence readings to an actual alert system (Chorus and Welker, 2021; Chorus and Bartram, 1999). However, the lack of regulations or guidelines for this parameter makes its use and understanding challenging for water managers, although its integration has been shown to be feasible (Ahn and Oh, 2007) and our results corroborate so.

Regarding model performances, LR models had the best results in terms of the studied metrics for short forecasting horizons until the 16th day mark. This performance, however, was much lower for longer forecasting horizons due to its reactive nature. Starting with the 16th day mark, the multivariate LSTM model outperformed the rest of the studied models in all studied metrics. Its performance is especially noteworthy in terms of the skill metric, the forecasting specific metric introduced in this paper, achieving almost a 25 % positive skill for the longest horizons, while none of the other models achieved even marginal positive skill. This positive value could be interpreted as the model capability to make real predictions beyond a purely reactive model. It is remarkable that the LR and RF were strongly affected by the increased input time of the variables, especially the LR (thus, they were limited to data from the previous 12 days), whereas the LSTM model was able to handle and take advantage of this additional data and complexity. On the other hand, the LSTM model demonstrated high accuracy and robustness, but this performance came at the cost of high sensitivity to changes in its configuration. In contrast, the LR and RF models showed less sensitivity to configuration changes. Due to LSTM success, in addition to the proposed comparison with the included models we propose the introduction of future works where LSTM is compared with other deep learning methods including all the relevant models for sequence forecasting (e.g., CNN, LSTM, GRU, RNN, autoencoders, GAN…). Traditional evaluation metrics had clear limitations to evaluate the forecasting ability of the models. These metrics alone did not provide enough information about the true predictive capability of the models and a comparison with purely reactive models, as in the skill metric, must be introduced. Limitations of these traditional metrics have been commonly discussed in the literature (Alexander et al., 2015; Hyndman and Koehler, 2006). In addition to this, there are also other metrics that measure important aspects of cyanobacterial bloom forecasting that were not included in this paper, remaining beyond of its scope. For example, those that quantitatively measure the fit of the predictions to the magnitude, timing, and shape of the relevant peaks (Donaldson et al., 2023), beyond the qualitative assessment made in this work in this regard. Future work on prediction applications using these metrics, or any prediction-specific metric (e.g., the skill metric proposed in this paper) is encouraged and recommended. In any case, the choice of the selected evaluation method must be focused on the specific objectives of the forecasting task (Dietze et al., 2018) and should be justified for each application. In this paper, a hybrid evaluation system that combines regression and classification metrics with a critical discussion of the forecasts at different forecasting horizons was proposed. Testing different forecasting horizons assisted to clearly identify the performance, strength and weakness of each model as these became more prominent at extended forecasting horizons (Thomas et al., 2018).

Forecasting PC concentration has great potential to enable timely action to minimize the impacts derived from cyanobacterial blooms (Bertone et al., 2018; Rousso et al., 2020). Although progress is being made, making environmental forecasts useful to decision makers is yet a challenge. This challenge is especially important in highly institutionalized sectors of critical resources, such as water management, where considerable efforts are still needed (Feldman and Ingram, 2009; Semmendinger et al., 2022). While defining the multiple sources of uncertainty is key in this process, forecasting limitations must be considered and never underestimated especially when management measures have to be taken (Wynne, 1992). Some examples of these limitations in our work begin with the inherent nature of the data source, such as the bias in fluorometric readings, or the fact that the data come from a profiler located at a specific site, which limits its ability to represent the entire water system. Some of these limitations could be counteracted by combining fluorometric readings of PC profilers with other tools such as remote sensing via satellites or drones providing surface spatial information, but these technologies require still more development (Almuhtaram et al., 2021a; Bertone et al., 2018; Rousso et al., 2020). However, current EWS, such as the one proposed in this paper, can already be very useful in promoting additional cost-effective verifications, such as simple visual inspections by operators, or increased attention while keeping an eye on the evolution of the alerts. Any more drastic management measures triggered by a forecasting alert system of this type should undergo previous field verification and expert judgment (Dietze et al., 2018). Taking into account these considerations, we propose the following ideas for future applications:

- Defining and addressing the gaps to connect cyanobacterial bloom forecasts with environmental decision-making.
- Exploring the potential of combining automated sensors mounted on profilers with other tools such as remote sensing to spread the scope of the forecasts.
- Including more specific parameters that are easily measured through automatic sensors, such as PC fluorescence, into the legislation within the context of cyanobacterial blooms.
- Studying the introduction of other deep learning models for sequence forecasting within this context.
- Defining and studying specific metrics that are selected for the specific needs of the forecasting application.

## 5. Conclusions

In this paper, a framework for effective cyanobacterial bloom forecasting is proposed. Its first part consists of an exploration and pre-processing for the generation of time series for cyanobacterial bloom forecasting using incomplete high frequency data from a multi-parametric probe at multiple depths. Using this method two different sets of time series were originally proposed, one intended to represent extreme values (time series of maximum values) and another to represent the integration of the water column (time series of mean values). Due to the high correlations between the maximum and integrated time series, the high amount of interpolated data, and the significant smoothing of information during critical periods, time series of integrated values were excluded from the modelling. Thus, six different models were proposed and studied including the autoregressive and multivariate versions of LR, RF, and the LSTM neural network. To ensure comparability, all of them had access to the same data set, but for each of them, optimal configurations were chosen in terms of input and parameters considered. The resulting forecasts were studied and compared using traditional regression and classification metrics as well as forecasting-specific metrics. This helped to assess how well the models could predict PC concentration and the proposed alert level. For short forecast horizons (i.e., <2 weeks), highly reactive models such as LR performed better according to the metrics, but they lacked anticipatory capability. For longer forecasting horizons (i.e., >2 weeks), LSTM models provided the best performance in all the evaluations performed. Moreover, LSTM model provided outstanding results when using PC and TEMP variables as inputs showing consistent results across all forecasting horizons, achieving an $R^2$ greater than 0.6 for PC concentration prediction, accuracies up to 0.9 for the alarm level prediction, and positive skills up to 25 % for the longest horizons.

## CRediT authorship contribution statement

**Claudia Fournier:** Writing – review & editing, Writing – original draft, Visualization, Validation, Supervision, Software, Resources, Project administration, Methodology, Investigation, Formal analysis, Data curation, Conceptualization. **Raúl Fernandez-Fernandez:** Writing – review & editing, Writing – original draft, Visualization, Validation, Supervision, Software, Methodology, Investigation, Formal analysis, Data curation, Conceptualization. **Samuel Cirés:** Writing – review & editing, Supervision, Project administration, Methodology, Investigation, Funding acquisition, Conceptualization. **José A. López-Orozco:** Writing – review & editing, Supervision, Resources, Project administration, Methodology, Funding acquisition, Conceptualization. **Eva Besada-Portas:** Writing – review & editing, Supervision, Resources, Project administration, Methodology, Funding acquisition, Conceptualization. **Antonio Quesada:** Writing – review & editing, Supervision, Resources, Project administration, Methodology, Investigation, Funding acquisition, Conceptualization.

## Declaration of competing interest

The authors declare that they have no known competing financial interests or personal relationships that could have appeared to influence the work reported in this paper

## Data availability

Data will be made available on request.

## Acknowledgements

This work was funded by the EU and the Spanish Ministry of Science and Innovation through the AIHABs consortium, supported by the ERA-NET AquaticPollutants Joint Transnational Call (GA Nº 869178), and by the Agencia Estatal de Investigación (Spain) through the grant PCI2021-121915. Additionally, support was provided by the Research Projects IA-GES BLOOM-CM (Y2020/TCS-6420) of the Synergic program of the Comunidad Autónoma de Madrid, SMART-BLOOMS (TED2021 130123B-100) funded by MCIN/AEI/10.13039/501100011033 and the European Union Next Generation EU/PRTR, and INSERTION (PID2021-127648OB-C33) of the Knowledge Generation Projects program of the Spanish Ministry of Science and Innovation. Special thanks to Dr. Ana Justel for her valuable insights during the conceptualization of this work. We also thank *Ecohydros* and Dr. Agustín P. Monteoliva for providing access to the floating platform data and cyanobacterial bloom-related information.

## Supplementary materials

Supplementary material associated with this article can be found, in the online version, at doi:10.1016/j.watres.2024.122553.